\documentstyle[aps,prl,twocolumn,epsfig]{revtex} 
\def\lsim{\lower.8ex\hbox{$\buildrel<\over\sim$}} 
\def\gsim{\lower.8ex\hbox{$\buildrel>\over\sim$}} 
\begin{document} 
 
%\draft 
\preprint{\today} 
\twocolumn[\hsize\textwidth\columnwidth\hsize\csname@twocolumnfalse\endcsname 
\title{A Hydrodynamic model for a dynamical jammed-to-flowing transition in 
gravity driven granular media}

\author{ 
Lyd\'eric Bocquet$^{(1)}$ Jalal Errami$^{(1)}$ and T. C. 
Lubensky$^{(2)}$} 
\address{(1) Laboratoire de Physique (UMR CNRS 5672)\\ 
ENS-Lyon, 46 all\'ee d'Italie, 69364 Lyon Cedex 07, France\\ 
(2) Physics Department, University of Pennsylvania, Philadelphia, Pennsylvania 19104, USA}

\date{\today} 
 
\maketitle 
 
\begin{abstract} 
Granular material on an inclined plane will flow like a fluid if 
the angle $\theta$ the plane makes with the horizontal is large 
enough. We employ a modification of a hydrodynamic model 
introduced previously to describe Couette flow experiments to 
describe chute flow down a plane. In this geometry, our model 
predicts a jammed-to-flowing transition as $\theta$ is increased 
even though it does not include solid friction, which might seem 
necessary to stabilize a state without flow. The transition is 
driven by coupling between mean and fluctuating velocity. In 
agreement with experiments and simulations, it predicts flow for 
layers with a thickness $H$ larger than a critical value $H_{\rm 
stop}(\theta)$ and absence of flow for $H<H_{\rm stop}(\theta )$. 
 
\end{abstract} 
 
\vskip2pc] 
 
In recent years, considerable experimental and theoretical effort 
has been devoted to the study of flow properties of granular 
media, yet no fully satisfying phenomenology of granular rheology 
has emerged \cite{Clement}.  The major difficulty in describing 
granular flows, particularly in dense media, is that both kinetic 
phenomena associated with particle motion and more static 
phenomena like solid friction and force chains are expected to be 
important. Following early work of Jenkins and Savage and of 
Haft\cite{Savage}, various kinetic-hydrodynamic threories based on 
coupled equations for the transport of mementum and kinetic energy 
have been constructed.  They assume instantaneous binary 
collisions and neglect any solid-like friction between grains. 
Various authors have proposed including shear independent terms in 
the stress tensor to capture empirically the effects of intergrain 
friction\cite{Johnson,Aronson}.  Neither of these types of 
theories incorporates the effects of strongly inhomogenoeus force 
networks in static and slowly driven dense granular 
materials\cite{Behringer,Mills} 
 
On the basis of experimental results on sheared granular material 
in Couette geometry, we recently proposed a modified hydrodynamic 
description\cite{Losert}, whose aim is to include high-density 
effects in equations describing energy and momentum transport. 
Specifically, in analogy with the behavior of supercooled liquids 
close to the glass transition, we assumed an ``anomalous'' 
divergence of the viscosity close to random close packing. Even 
though our approach still does not include enduring contacts and 
solid friction between grains, it is able to reproduce most of the 
experimental results in the Couette Geometry. 
 
Granular flow down inclined slopes is of considerable practical 
importance in geophysical phenomena such as rock and snow 
avalanches, pyroclastic flows, etc..  Systematic 
experiments\cite{Pouliquen} and molecular simulations\cite{Grest} 
on a layer of beads on an inclined plane as a function of layer 
thickness $H$ and angle $\theta$ of the plane relative to 
horizontal yield a curve $H_{\rm stop} ( \theta )$ separating 
jammed, stationary behavior at small $H$ from flowing behavior at 
large $H$.  Moreover, precise and systematic measurements of the 
mean-bead velocity yield a maximum velocity that scales as 
$H^{3/2}$. 
 
In this paper, we analyze chute flow down inclined planes using 
the frictionless kinetic-hydrodynamic theory developed for the 
study of Couette flows\cite{Losert}.  Our analysis produces 
results in agreement with experiment and simulation in this 
geometry and provides further evidence that our phenomenological 
kinetic-hydrodynamic theory provides a robust and general 
description of flowing granular media.  In particular, we find an 
$h_{\rm stop}$ curve, or equivalently a critical angle separating 
stationary from flowing behavior. Normally, the existence of such 
a critical angle $\theta_c$ is associated with solid friction: a 
solid block on an inclined plane is stationary for 
$\theta<\theta_c$ and slides for $\theta>\theta_c$.  In our 
analysis, the existence of $\theta_c$ is associated with a purely 
dynamical jammed-to-flowing transition in which solid friction is 
totally absent.  The origin of this transition is the coupling 
between velocity fluctuations and mean flow - an important feature 
of our previous study. 
 
%\narrowtext 
{\bf Geometry of the problem}. We consider the geometry of the 
experiments in Refs. \cite{Pouliquen}: a layer of granular 
material, composed of spheres of diameter $d$, of thickness $H$ 
flows on an inclined plane making an angle $\theta$ with the 
horizontal. We choose a coordinate system with the $x$-axis 
parallel to the plane, the $z$-axis perpendicular to it, and 
origin at the bottom of the flowing layer. 
%, as depicted in Fig \ref{fig0}. 
%\begin{figure} 
%\psfig{file=./fig0Bocquet.eps,width=7.5cm,height=5.0cm} 
%\caption{Geometry of the system.} 
%\label{fig0} 
%\end{figure} 
In a stationary state, conservation of momentum in directions 
parallel and perpendicular to the plane leads to the equations: 
$\partial_z \sigma_{xz}= \rho(z) g \sin \theta$ and $\partial_z 
\sigma_{zz}= \rho(z) g \cos \theta$, where $\rho$ is the mass 
density and $\sigma_{zz}$ and $\sigma_{xz}$ are, respectively, the 
diagonal and off diagonal components of the stress tensor. 
Combining these two model-independent equations yields the simple 
relation: $\sigma_{xz}=\tan \theta \sigma_{zz} + C$, where $C$ is 
a constant independent of $z$. In our hydrodynamic model, the off 
diagonal component of the stress tensor is {\bf viscous like}: 
$\sigma_{xz}=\eta (\rho,T) \dot\gamma$, where $\dot\gamma=dV_x/dz$ 
is the shear rate, $V_x$ the mean velocity, and $\eta$ is the 
viscosity, a function of the density and granular temperature 
field $T$.  As usual the latter is defined in terms of the RMS 
part of the velocity field $T(z)={1\over 3} m \langle v^2(z) 
\rangle$. The temperature profile is determined by the balance 
between viscous heating, heat flow, and dissipation through 
inelastic collisions \cite{Losert}: $ {\partial \over {\partial 
z}} \lambda(\rho,T){\partial \over {\partial z}} T +\sigma_{xz} 
\dot \gamma - \epsilon (\rho,T) T =0$, where $\lambda(\rho,T)$ 
is the thermal conductivity and $\epsilon (\rho,T)$ is the rate of 
kinetic energy loss. These hydrodynamic equations are closed by 
the equation of state $P=f(\rho) T$. In contrast to classical 
thermalized fluids, there are strong temperature and density 
variations over the system, and the density and temperature 
dependence of transport coefficients play an important role in 
determining flow properties. These dependences are usually 
obtained using kinetic theory \cite{Savage} with the Enskog 
approximation for the collision kernel, which accounts for 
excluded volume but neglects any correlation between collisions. 
By construction this theory yields transport coefficients that are 
roughly proportional to the collision frequency, {\it i.e.}, to 
the pair-correlation function at contact, which in the 
high-density limit is expected to diverge like $(\rho_c - 
\rho)^{-1}$ close to the maximum density, $\rho_c$, allowed by 
excluded volume effects \cite{Speedy}. The assumption of 
negligible correlation between collisions should fail at high 
density, and in our modified description of Ref. \cite{Losert}, we 
proposed replacing the Enskog expression for the viscosity with 
one with a specific ``anomalous" algebraic divergence, $\eta \sim 
(\rho_c-\rho)^{-\beta}$ \cite{note} close to $\rho_c$, in full 
analogy with the behavior of supercooled liquids close to the 
glass transition \cite{Gotze}. This picture is supported by the 
experimental observation in sheared granular material of a 
specific scaling law relating the local shear rate $\dot \gamma$ 
to the granular temperature $T$ \cite{Losert,Mueth}. 
 
In the Couette geomety, the density is high throughout the cell. 
In contrast, in the current chute geometry, small densities could 
in principle be reached at the top surface, and we need to include 
the low-density limit in our expressions for transport 
coefficients. Since our aim is to provide a sketch of the 
jammed-to-flowing, we shall not use the full Enskog expression for 
transport coefficient in the small and intermediate density 
regions, as is done by various authors \cite{Savage}. Instead, we 
will use simple interpolated expressions that are consistent with 
both the high-, intermediate-, and small-density limit. Our 
specific choices are $\lambda(\rho,T)= [\lambda_0/(m^{1/2} d^2)] 
(1-\rho/\rho_c)^{-1} T^{1/2} $, $\epsilon(\rho,T)= 
[\epsilon_0/(m^{1/2} d)] \rho (1-\rho/\rho_c)^{-1} T^{1/2} $, 
$\eta(\rho,T)=\eta_0 [m^{1/2}/d^2] g(\rho) T^{1/2} $, where $m$ 
the mass of one particle, $d$ is its diameter, and $\lambda_0$, 
$\epsilon_0$ and $\eta_0$ are dimensionless constants. The 
function $g(\rho)$ is chosen so as to interpolate between the very 
low-density limit, where the viscosity goes to a constant, to its 
high-density ``anomalously" diverging behavior close to $\rho_c$, 
passing through its Enskog behavior for intermediate (but already 
high) densities. A sensible choice is simply 
$g(\rho)=(\alpha(\rho)^{\beta-1}+\alpha_0^{\beta-1}) 
\alpha(\rho)^{-\beta}/(1+\alpha_0^{\beta-1})$, where 
$\alpha(\rho)=1-\rho/\rho_c$ and $\beta$ is the anomalous exponent 
discussed above. The parameter $\alpha_0$ fixes the density where 
the crossover to the anomalous regime occurs. The equation of 
state is written as $P/T=\rho/(1-\rho/\rho_c)$, which does indeed 
reduce to the correct limits in the low- and high-density 
limits\cite{Speedy}. We emphasize that our results are not very 
sensitive to the specifics forms for $g$ and the equation of 
state. Our expressions were chosen to minimize the number of 
parameters in the model. 
 
The previous equations can be made dimensionless using a reduced 
velocity $v=V/\sqrt (g d)$, temperature $t=T/(mg d)$, stress 
$\sigma=\sigma_{xz}/(\rho_c g d)$, pressure $p=P/(\rho_c g d)$, 
and distance $\tilde z=z/d$ (we shall omit the tilde over $\tilde 
z$ in the following). Using the previously defined expressions for 
the transport coefficients, we can now rewrite the coupled 
equations for stress and temperature as: 
\begin{mathletters} 
\begin{eqnarray} 
&{\partial \over {\partial z}} p = -\cos \theta {t\over p+t} 
\label{hydroa} \\ 
&\partial_z \left( \left({p+t}\over t^{1/2}\right) \partial_z t 
\right) + a  {\sigma^2 \over g({p\over p+t}) t^{1/2}} - b p 
t^{1/2} = 0 , 
\label{hydrob} 
\end{eqnarray} 
\label{hydro} 
\end{mathletters} 
where $p/(p+t)=\rho/\rho_c$ is the reduced density, $a=(\rho_c 
d^3)^2 /(\lambda_0 \eta_0)$ and $b=\epsilon_0 \rho_c 
d^3/\lambda_0$ are dimensionless constants, $\sigma$ is related to 
$p$ according to $\sigma=\tan \theta p +c$, with $c$ an as yet 
unspecified constant. The parameters appearing in Eq.\ 
(\ref{hydro}) are strongly constrained by the complementary 
experiments performed in the Couette cell \cite{Losert,Mueth}. In 
this case, the velocity and temperature profiles are localized 
close to the moving boundary over a depth of a few particles 
diameter. This behavior is predicted by the hydrodynamic model 
with a characteristic decay lentgh for the RMS velocity given by 
$\delta/d=(2/b)^{1/2}$ \cite{Losert}. Experimentally $\delta$ is 
of the order of 4-5 particle diameters, yielding $b \simeq 0.1$ 
(we shall use in the following $b=0.111$). The exponent $\beta$ 
was determined experimentally in Ref.\ \cite{Losert} to be 1.75, 
while experiments over a larger range of data by Mueth yielded 
$\beta=1.5$. The parameter $\alpha_0$ is chosen such that the 
cross over to the anomalous scaling occurs roughly 10\% below 
random close packing. In pratice $\alpha_0=0.05$ is adequate. 
Eventually, $a$ is expected to be of order unity. 
 
The hydrodynamic equations (\ref{hydro}) have to be supplemented 
by boundary conditions for the velocity and temperature fields at 
the bottom and top surfaces. As to the conditions on the granular 
temperature, we showed in our previous study of the Couette cell 
that a vanishing ``heat flux" condition at the boundaries yielded 
results in agreement with the experimental ones. In the present 
study, we shall thus assume $\partial_z T=0$ at the boundaries. In 
the recent simulations of Silbert et al. \cite{Grest}, the 
temperature profile has a vanishing derivative close to the bottom 
and top boundaries. The previous assumption thus amounts to 
slightly shifting the hydrodynamic boundaries of the system, a 
behavior which is encountered in classical fluids \cite{BB}. The 
boundary condition for the mean velocity field are {\it a priori} 
more obvious: at the wall surface, the velocity is assumed to 
vanish, while at the top surface a stress-free boundary condition 
will be assumed, $\sigma=0$, yielding $\partial_z v=0$ at $z=H$. 
Since, as mentioned above, the effective hydrodynamic boundary 
$z=H$ is located slightly inside the material, the pressure is not 
expected to vanish there, and it can be written as $p_0=z_0 \cos 
\theta$ (in reduced units) with $z_0$ a {\it molecular} distance 
of the order of the diameter (unity in reduced units). To be fully 
consistent with the stress free boundary for the velocity, we thus 
write $\sigma=\tan \theta (p-p_0)$ ({\it i.e.}, fix the constant 
$c=-\tan \theta p_0$). As emphasized above, this is a molecular 
effect that disappears in the large-$H$ limit. Moreover this 
choice does not affect the existence of the jammed-to-flowing 
transition to be discussed now.

\begin{figure} 
\psfig{file=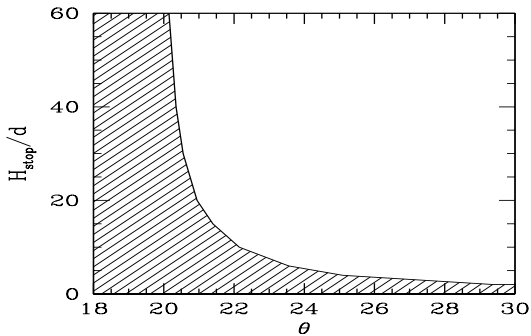,width=7.5cm,height=5.0cm} 
\caption{Phase diagram for the jammed-to-flowing transition in a 
gravity driven granular material as obtained from the resolution 
of the hydrodynamic equations (\ref{hydro}). The parameters used 
are $a=1.3$, $b=0.111$, $\alpha_0=0.05$, $z_0=0.5$.} 
\label{fig1} 
\end{figure}

The previous system of equations (\ref{hydro}) has been solved 
using a standard Runge Kutta algorithm, using a shooting-like 
procedure: for a given temperature $T(H)$ at the top boundary, 
the derivative of $T$ at the bottom, $\partial_z T (0)$ is 
computed. The solutions we seek thus correspond to zeros in the 
$\partial_z T (0)$ versus $T(H)$ curve.  A transition between two 
kinds of behaviour is then found: for a set \{$H,\theta$\} such 
that $H$ is below a critical $H_{\rm stop}(\theta)$ curve, only 
the $T=0$ solution exists. This corresponds to a ``jammed'' state, 
which stays at rest. Above this critical value, nonzero solutions 
for the temperature, which correspond to a flowing regime do 
exist. In Fig. \ref{fig1}, we plot the boundary between the two 
regimes, denoted $H_{\rm stop}(\theta)$ as in Ref. 
\cite{Pouliquen}. This curve is in qualitative agreement with the 
experimental results of Refs \cite{Pouliquen}. 
 
This shows that the hydrodynamic model is able to generate a 
finite critical angle below which no flow occurs. Within the 
model, the origin of this behavior is simply the balance between 
the ``viscous heating'', which generates fluctuations, and 
dissipation through collisions, which tends to inhibit the flow. 
The transition is thus purely dynamical: a jammed state occurs for 
low angles/thicknesses because fluctuations are insufficient to 
allow flow. 
 
\begin{figure} 
\psfig{file=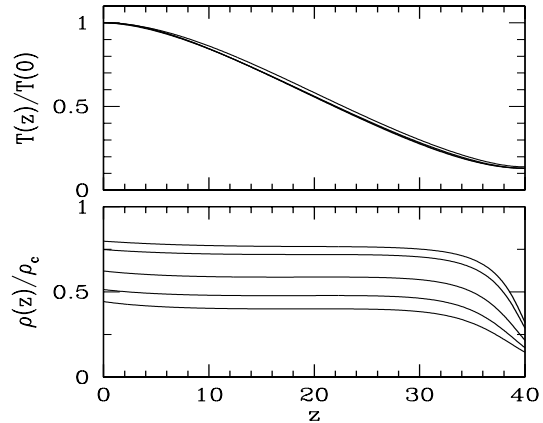,width=7.5cm,height=6.0cm} 
\caption{Top : Normalized temperature profiles for $H=40$, $\theta=20.55, 
20.90, 21.50, 22.50$. Bottom : Normalized density profiles for 
$H=40$, $\theta=20.55, 20.90, 22.50, 24.50, 26.50$ from top to bottom. 
($a=1.3$, $b=0.111$, $\alpha_0=0.05$, $z_0=0.5$)} 
\label{fig2} 
\end{figure}

Above $H_{\rm stop}(\theta)$, more than one solution can be found 
for some sets of parameters \{$H,\theta$\}. However a linear 
stability analysis of the hydrodynamic equation for the 
temperature shows that only a single nonzero solution is 
dynamically stable. On the other hand the $T=0$ solution is always 
linearly stable but becomes unstable against larger ``kicks'' in 
the temperature. It is important to note that there is a solution 
to the hydrodynamic equations for any set of $H$ and $\theta$ 
above the boundary $H_{\rm stop}(\theta)$, as is found 
experimentally. This is in contrast to other approaches where for 
a given $\theta$ a {\it single} flowing thickness $H$ is selected 
\cite{Johnson}. In the flowing regime, we obtain temperature and 
density profiles,which are plotted in Fig. \ref{fig2} for $H=40$. 
%\begin{figure} 
%\psfig{file=./fig2Bocquet.ps,width=7.5cm,height=5.0cm} 
%\caption{Temperature profiles for $H=40$, $\theta=$. The parameters used are 
%$a=1.3$, $b=0.111$, $\alpha_0=0.05$, $z_0=0.5$.} 
%\label{fig2} 
%\end{figure} 
%\begin{figure} 
%\psfig{file=./fig3Bocquet.ps,width=7.5cm,height=5.0cm} 
%\caption{Density profiles for $H=40$, $\theta=$. The parameters used are 
%$a=1.3$, $b=0.111$, $\alpha_0=0.05$, $z_0=0.5$.} 
%\label{fig3} 
%\end{figure} 
Both results for the density and temperature are in qualitative 
agreement with the simulations of Refs.  \cite{Grest}. In 
particular the density profile is found to be almost constant in 
the middle of the sample, with a large drop at the top surface, 
as found numerically. In the simulations however, the plateau in 
the density is flatter than ours. 
 
Velocity profiles are obtained by integrating the shear rate 
$\dot\gamma=dv_x/dz$, which in our model is linearly related to 
the non diagonal stress $\sigma_{xz}(z)=\eta (\rho(z),T(z)) 
\dot\gamma (z)$. We find that the $z$ dependence of the profiles 
is in very good agreement with the predictions of a simple Bagnold 
scaling $\sigma = A_{\rm Bag} \dot\gamma^2$, $v(z)/v_{\rm 
max}=1-(1-z/H)^{3/2}$. Within the Bagnold assumption, $v_{\rm max} 
= (2/3) A_{\rm Bag} \sqrt{\sin\theta}H^{3/2}$.  This agreement 
simply indicates that {\it once the system flows}, Bagnold-like 
scaling is a good approximation to the flow properties, even 
though {\it it cannot predict a jammed-to-flowing transition}. The 
same agreement with Bagnold scaling was observed in the 
simulations of Ref.\ \cite{Grest}.  The $H^{3/2}$ dependence of 
the maximum velocity $v_{\rm max}$ is also recovered in our model 
in agreement with experiments and simulations.  Our results are 
also consistent with a $\sqrt{\sin\theta}$ dependence of $v_{\rm 
max}$ with $A_{\rm Bag} \approx 0.2$.  This is in fair agreement 
with the simulations of Silbert {\it et al.} \cite{Grest}, who 
find a slight dependence of $A_{\rm Bag}$ on $\theta$.  On the 
other hand, the experiments of Pouliquen exhibit a stronger 
dependence on $\theta$, with $v_{\rm max}$ scaling like $[H_{\rm 
stop}( \theta)]^{-1}$ 
 
{\bf Discussion.} Reference \onlinecite{Losert} introduced a 
modified kinetic-hydrodynamic theory for granular flow that 
predicted temperature and velocity profiles in dense systems in 
Couette geometry in agreement with experiments.  In this paper, we 
calculate the corresponding profiles for flow down an inclined 
plane using a generalization of this theory to include gravity and 
low- as well as intermediate- and high-density regimes.  Our 
results agree with those obtained in experiments\cite{Pouliquen} 
and simulations\cite{Grest}.  Thus, the kinetic-hhydrodynamic 
theory of granular flow provides a good description of flow in two 
qualitatively different geometries. Though it is premature to 
generalize from two to all geometries, it would, nevertheless, 
appear that this description might serve as an all purpose 
phenomenological theory for granular flow.  In the case of flow 
down a plane, it is able, contrary to intuition, to predict a 
jammed-to-flowing transition and a critical angle below which no 
flow takes place, even though it does not include solid friction 
between grains, which might be viewed as necessary to stabilize a 
state without flow. 
 
Our results raise a number of basic questions about the 
mechanisms underlying granular flows.  Two important features of 
granular materials are (i) the possibility of solid friction 
between grains and (ii) the absence of velocity fluctuations 
({\it i.e., temperature}) when the system is at rest.  Under 
flow, the effects of these features become intertwined, making it 
difficult to provide first principles modeling of the rheology of 
granular materials.  The fact that our frictionless hydrodynamic 
model produces successful predictions for two different 
geometries suggests that solid friction is not necessarily a 
dominant feature (though it may not be negligible). On the other 
hand, our theory does highlight the importance of coupling between 
velocity fluctuations and mean flow.  Intuitively, one might 
expect this to be a generic feature of rheology of ``$T=0$" 
glassy systems:  in order to flow, a system must create its own 
fluctuations, which in turn couple to flow.  It is interesting to 
note that various descriptions based on similar ideas have 
recently emerged in attempts to provide a phenomenological 
description of the rheology of complex systems such as gels, 
paste\cite{Armand}, and foam\cite{Debregeas}.  The hydrodynamic 
model successfully captures coupling between mean and RMS velocity 
fields, and it should serve as a basis for developing improved 
models that could for example include contact forces.  At high 
density, enduring contacts occur, but they must not be confused 
with the existence of solid friction between grains. Indeed, 
enduring contacts do not invalidate a hydrodynamic description; 
they can be taken into account in transport equations (as done 
for example in simple liquids with continuous 
interactions\cite{JPH}).  In this case, only the $T^{1/2}$ 
scaling of transport coefficients, which stems from the 
assumption of binary collisions, will be modified, and a more 
complex temperature dependence of transport coefficients is 
expected. 
 
L.B. acknowledges fruitful discussions with O. Pouliquen, Y. 
Forterre and G. Debregeas.  T.C.L. acknowledges the support of the 
National Science Foundation under Grant DMR00-96532 and of the 
E.N.S. de Lyon, where this work was initiated.


\begin{references} 
 
\bibitem{Clement} E. Clement, Curr. Opinion in Coll. and Interface Sci. {\bf 4}, 
294 (1999). 
 
\bibitem{Savage} J.T. Jenkins and S.B. Savage, J. Fluid Mech. {\bf 130} 186 (1983); P.K. Haft, 
J. Fluid Mech. {\bf 134} 401 (1983) 
 
\bibitem{Johnson} P.C. Johnson, P. Nott and R. Jackson, J. Fluid. Mech. {\bf 210} 501 
(1991); K.G. Anderson and R. Jackson, J. Fluid Mech. {\bf 241}, 
145 (1992). 
 
\bibitem{Aronson} I.S. Aronson and L.S. Tsimring, Phys. Rev. E {\bf 64} 
020301 (2001). 
 
 
\bibitem{Behringer} H.M. Jaeger, S.R. Nagel and R.P. Behringer, Rev. Mod. Phys. {\bf 68}, 1259 (1996). 
 
\bibitem{Mills} P. Mills, D. Loggia, M. Tixier, Europhys. Lett. {\bf 45}, 733 (1999); G. Debregeas 
and C. Josserand, Europhys. Lett. {\bf 52}, 137 (2000). 
 
\bibitem{Losert} W. Losert, L. Bocquet, T.C. Lubensky and J.P. Gollub, Phys. Rev. Lett. {\bf 85} 1428 
(2000); L. Bocquet {\it et al.}, Phys. Rev. E, in press (2001). 
 
\bibitem{Pouliquen} O. Pouliquen, Physics of Fluids, {\bf 11} 542 (1999); 
A. Daerr and S. Douady, Nature {\bf 399} 241 (1999); E. Azanza, 
F. Chevoir, P. Moucheron, J. Fluid. Mech. {\bf 400} 199 (1999). 
 
\bibitem{Durian} P.A. Lemieux and D. Durian, Phys. Rev. Lett. {\bf 85}, 4179 (2001) 
 
 
\bibitem{Grest} D. Ertaz {\it et al.} Europhys. Lett. {\bf 56}, 214 (2001); L. Silbert {\it et al.} {\tt condmat}/0105071 
 
 
\bibitem{Speedy} R.J. Speedy, J. Chem. Phys. {\bf 110}, 4559 (1999). 
 
%\bibitem{Alder} B. Alder, D. Gass, T. Wainwright, J. Chem. Phys. {\bf 53}, 3813 (1970). 
 
\bibitem{Gotze} W. Goetze and L. Sjoegren, Rep. Prog. Phys.  {\bf 55}, 241 (1992). 
 
 
\bibitem{note} Note that this anomalous behavior is not expected to occur fot the the energy transport coefficients ($\lambda$ and $\epsilon$). 
%, because the latter are less affected by cage effects : their Enskog 
%expression 
%is thus expected to hold even at high densities. 
See Ref. \cite{Losert} for a full discussion concerning this point. 
 
\bibitem{BB} L. Bocquet and J.L. Barrat, Phys. Rev. E {\bf 49}, 3079 (1994). 
\bibitem{Mueth}  D. M\"uth, {\tt condmat}/0103557. 
 
\bibitem{JPH} J.P. Hansen and I.R. Mc Donald, {\it Theory of Simple Liquids}, 2nd ed. (Academic Press, London, 1986). 
 
\bibitem{Armand} C. Derec, A. Ajdari, F. Lequeux, Eur. Phys. Jour. E {\bf 4}, 355 (2001). 
 
\bibitem{Debregeas} G. Debr\'{e}geas, H. Tabuteau, J.M. di Meglio, 
Phys. Rev. Lett. {\bf 87}, 8305 (2001). 
 
\end{references}
\end{document}